\def\year{2022}\relax
\documentclass[letterpaper]{article} 
\usepackage{aaai22}  
\usepackage{times}  
\usepackage{helvet}  
\usepackage{courier}  
\usepackage[hyphens]{url}  
\usepackage{graphicx} 
\usepackage{natbib}  
\usepackage{caption} 
\DeclareCaptionStyle{ruled}{labelfont=normalfont,labelsep=colon,strut=off} 
\usepackage{algorithm}
\usepackage{algorithmic}

\usepackage{newfloat}
\usepackage{listings}
\floatstyle{ruled}
\newfloat{listing}{tb}{lst}{}
\floatname{listing}{Listing}
\usepackage{amsfonts}
\usepackage{amssymb}
\usepackage{amsmath}
\usepackage{graphicx}
\usepackage{bm}
\usepackage{booktabs}
\usepackage{mathtools}
\usepackage{caption}
\usepackage{subcaption}
\usepackage{bbding}
\usepackage{pifont}

\title{NewsQuote: A Dataset Built on Quote Extraction and Attribution for Expert Recommendation in Fact-Checking}
\author {
   Wenjia Zhang,\textsuperscript{\rm 1}
   Lin Gui, \textsuperscript{\rm 2}
   Rob Procter, \textsuperscript{\rm 1,3}
   Yulan He \textsuperscript{\rm 1,2,3}
}
\affiliations {
   \textsuperscript{\rm 1} Department of Computer Science, University of Warwick, UK\\
   \textsuperscript{\rm 2} Department of Informatics,King's College London, UK\\
    \textsuperscript{\rm 3} The Alan Turing Institute, UK\\
   \{Wenjia.Zhang,Rob.Procter\}@warwick.ac.uk, \{lin.1.gui,yulan.he\}@kcl.ac.uk
}
\usepackage{bibentry}

\begin{document}

\maketitle

\begin{abstract}
To enhance the ability to find credible evidence in news articles, we propose a novel task of expert recommendation, which aims to identify trustworthy experts on a specific news topic. To achieve the aim, we describe the construction of a novel NewsQuote dataset consisting of 24,031 quote-speaker pairs that appeared on a COVID-19 news corpus. We demonstrate an automatic pipeline for speaker and quote extraction via a BERT-based Question Answering model. Then, we formulate expert recommendations as document retrieval task by retrieving relevant quotes first as an intermediate step for expert identification, and expert retrieval by directly retrieving sources based on the probability of a query conditional on a candidate expert. Experimental results on NewsQuote show that document retrieval is more effective in identifying relevant experts for a given news topic compared to expert retrieval.\footnote{Our source code can be accessed at: \url{https://github.com/WenjiaZh/NewsQuote}}

\end{abstract}

\section{Introduction}
The rapid growth of misinformation in recent years has been the subject of much attention from academia, journalists, political analysts and fact-checking organisations and has prompted research into NLP-based techniques and tools to support fact-checking work and evidence verification \cite{lazarski2021using, zeng2021automated, guo2022survey}. Much of this research effort has been based on a \textit{document-centric} model of fact-checking work, where the end goal is to provide the journalist or fact-checker with an (automated) ranked list of documents relevant to the claim that they can then use as evidence for determining its likely veracity (e.g., \citet{zhao2023panacea}). 

Our recent research reveals that some fact-checkers use a \textit{expert-centric} model, whereby they search for credible and trustworthy experts who are willing to be quoted \cite{procter2023}. 
Finding such experts is a big challenge and often journalists and fact-checkers aim to interview several experts as relying solely on one source may not be considered as sufficiently credible. In the case of contentious claims, they may also need to ensure their reports are balanced \cite{procter2023}.

There is thus an urgent need to develop a tool for journalists and fact-checkers to search for experts based on their past record of being quoted by news media and fact-checking organisations, and other trustworthy agencies. To achieve this goal, we need to first automatically extract quotes and their sources from news articles, and then second return a ranked list of experts relevant to a query that then can be assessed by the journalist or fact-checker. This can be formulated as two tasks: (1) quote extraction and attribution, and (2) expert recommendation.

For the first task of quote extraction and attribution, most datasets were built on literature narratives and limited in size due to the reliance on manual annotation \cite{zhang2003identifying,elson2010automatic,fernandes2011quotation,lee-yeung-2016-annotated}. But newswire has much fewer monologues and dialogues than fiction \cite{o2012sequence}. Early work relied on rule-driven frameworks and manually-defined linguistic patterns, hence they mainly focused on direct quotes \cite{lee-yeung-2016-annotated,zhang2021directquote,vaucher2021quotebank}.
Unlike play scripts or fiction, people quoted in the news media are not limited to a list of fixed characters. In addition, the constantly evolving stream of events reported in news articles and diverse writing styles used by news media outlets make it difficult to identify experts and extract quotes by relying on regular expressions. 

For the second task of expert recommendation, much work has been conducted for expert finding in academic research \cite{sun2015leverage,silva2014research,wang2017context}, online communities \cite{yuan2020expert}, and the enterprise field \cite{paul2016find,askari2022expert}. However, we are not aware of any work searching for experts based on their track record of being quoted in news articles. 
\begin{table*}
\centering
\begin{tabular}{ccccl}
\toprule
\textbf{Corpus}&\textbf{\#Quotes}&\textbf{Indirect\%}&\textbf{Entity}&\textbf{Data Source}\\
\midrule
StylisticsCorpus & 16,533  & 16  & \ding{55} & Fiction, Newspaper, Biographies\\
\midrule
PARC3 & 19,712  & 72  & \ding{55} & Wall Street Journal \\
\midrule
QuoteBank & 178 million  &  - & \ding{51} & News Articles\\
\midrule 
DirectQuote & 10,279 &  0 & \ding{51} & News Articles\\
\midrule
\textsc{NewsQuote} & 24,031 & 81 & \ding{51} & News Articles\\
\toprule
\end{tabular}
\caption{Summary of large-scale (larger than 10,000)  news-originated English quotation corpora.}
\label{tab:datas}
\end{table*}
In this paper, we propose a semi-automatic approach to construct a news quotation dataset, called \textsc{NewsQuote}, from the AYLIEN coronavirus dataset\footnote{This data was aggregated, analyzed, and enriched by AYLIEN using the AYLIEN’s News Intelligence Platform. \url{https://aylien.com/resources/datasets/coronavirus-dataset},\url{https://aylien.com/blog/free-coronavirus-news-dataset}}, which contains over 1.5 million English news articles generated from around 440 global sources. We utilise the semantic role labelling results of sentences in news articles to extract the quote trigger verbs, subjects (i.e., sources) and objects (i.e., quotes), and identify sources by their corresponding DBpedia\footnote{\url{https://www.dbpedia.org/}} ontology class labels. The resulting dataset contains both direct and indirect quotes, and also mixed quotes where only part of the quotations is placed inside quotation marks. We introduce the task of finding sources of evidence from news reports and present a set of approaches for (1) identifying quotations and their sources from text; and (2) recommending potential experts for given news topics. Our experimental results illustrate the feasibility of using our constructed \textsc{NewsQuote} dataset for developing an automated tool for searching and ranking subject-matter experts for journalists and fact-checkers.

\section{Related Work}
\paragraph{Quotation Extraction and Attribution}
Quotation extraction and attribution originated as a study of literary works \cite{zhang2003identifying}, and now typically covers three sub-tasks: identifying sources, extracting quotations, and attributing a quotation to its source.
In Table \ref{tab:datas}, we summarise several large-scale English quotations datasets that are built on news articles. 
The \textbf{StylisticsCorpus} \cite{semino2004corpus} was designed for discourse presentation in written British narratives. They opted for hard news (e.g., accidents, conflicts, and crimes) \cite{bell1991language} as a part of the data source because of its circulation, narrative, authenticity, and cultural prominence. Of the total data, 5407 occurrences came from the press. They classified these samples into speech, writing, and thought. Then they divided each class into many presentation categories, such as indirect, free indirect, direct, and free direct.  
The \textbf{PARC3} \cite{pareti2016parc} project aims to fill the gap of the attribution relation (AR). Their annotation scheme tagged three constitutive components of an AR: source, cue, and content. They labeled the quote status as direct, indirect, or mixed by the usage of quote marks, and looked into the depth of attribution by the level of nesting. The inspiration for generating \textbf{QuoteBank} \cite{vaucher2021quotebank} came from the tangled nature of contemporary news flow. \citet{vaucher2021quotebank} exploited duplicate reports in different media to learn the patterns of quote-source pairs. Focusing on the attribution of direct quotations, they proposed an end-to-end minimally supervised framework, named Quobert, to extract and attribute quotations. Using Quobert, they generated QuoteBank from the Spinn3r dataset \cite{burton2011icwsm}, and linked source entities to the Wikidata knowledge base.
\textbf{DirectQuote} \cite{zhang2021directquote} contains direct quotations manually annotated from online news media. Like QuoteBank, each source can be linked to a Wikidata named entity to benefit various downstream tasks. 

Among the existing news quotation datasets, StylisticCorpus and PARC3 contain both direct and indirect quotes, but do not originate from multi-platform news stories, nor do they provide source-entity linking to Wikidata. The other two datasets, QuoteBank and DirectQuote, have each of their sources linked to a Wikidata named entity, but they only focus on direct quotes. In comparison, our \textsc{NewsQuote} contains various types of quotes including direct, indirect and mixed quotes where only part of the quotation is inside the quotation marks. In addition, all sources have their DBpedia entity links.

\paragraph{Expert Finding}

The core task in expert finding is to identify candidates with the required expertise for a given query \cite{yuan2020expert}. Therefore, solutions focus on matching the demand of searchers and the experience of relevant experts. In practise, this problem has expanded to different situations where various factors were considered.
Academic 
accounts for up to 65\% expert finding research \cite{husain2019expert}. When looking for academic experts, attention is given to topic relevance, expert quality, research connectivity (\citealp{sun2015leverage,silva2014research,wang2017context}), as well as capacity limitation \citep{neshati2014expert}. Meanwhile, many expert finding systems are used on online platforms, such as community question answering, social networks and forums \cite{yuan2020expert,faisal2017expert}. 
In the enterprise field, experts' accessibility and productivity are considered to have significant economic benefits (\citealp{silva2013social,paul2016find}). 
In the medical domain, when looking for the most suitable doctor for a particular patient, the patient's underlying conditions are of critical importance (\citealp{tekin2014discover}). In lawyer finding, users may prefer candidates in the same state or city, hence the physical location was emphasized \cite{askari2022expert}. 

\section{NewsQuote: Dataset Construction}


In this section, we describe how we constructed the dataset, including details of the data source, pre-processing steps performed, and test set annotation. Example data entries and dataset statistics will be presented at the end. 

\subsection{Data Collection}

We built our \textsc{NewsQuote} dataset from the AYLIEN coronavirus dataset, 
published between November 2019 and August 2020. We used the AYLIEN News API\footnote{\url{https://aylien.com/product/news-api}} to retrieve news articles. Apart from text, each article is also accompanied with the meta data such as authors, 
keywords, 
summary, source, publishing time, 
topical categories coded by both the Interactive Advertising Bureau (IAB) taxonomy\footnote{\url{https://www.iab.com}} and the IPTC NewsCodes\footnote{\url{https://iptc.org/standards/newscodes/}}, as well as recognized entities and entity links from  DBpedia.
\subsection{Pre-processing}
\paragraph{Data De-duplication}
As the same news story may be posted by multiple sources and there were exact duplicates in the original dataset, we removed news articles that are similar to ones already been published. 
News articles were first sorted in chronological order. News duplicates were then detected using a RoBERTa classifier\footnote{\url{https://huggingface.co/vslaykovsky/roberta-news-duplicates}} trained with title-body pairs using semi-supervised learning \cite{ruckle-etal-2019-neural}. For processing efficiency, the dataset was split into 16 equal-sized subsets. For each subset, titles and first sentence of news summaries of temporally-ordered news articles were sequentially fed as input to the RoBERTa classifier. Any duplicates were removed. 
After data de-duplication, 158,325 news articles remained. The total number of source platforms is 258, and as shown in Figure \ref{AC2}, the top 5 source platforms are: Daily Mail, Yahoo, Seeking Alpha, Business Insider, Reuters.

\paragraph{Quote Trigger Word Filtering}
For each of the selected articles, we segment the the main body into sentences, and then use a pre-trained BERT-based semantic role labeling model \cite{shi2019simple} to extract verbs (or predicates), subjects, and objects. We obtained a candidate verb list sorted by their occurrence frequencies. After manually checking the candidate verbs with over 100 occurrences, we identified 352 quote trigger words that are more likely indicative of direct or indirect quotes. 
The list of verbs are presented in our source code repository \footnote{\url{https://github.com/WenjiaZh/NewsQuote/blob/main/SelectedTriggerVerbs.csv}}. 
Some of the verbs are clearly indicative of quotes, such as `\emph{said}', while others may not be associated with quotes in a traditional sense, for example, `\emph{tweet}'. After identifying the quote trigger words, we only kept the sentences with at least one trigger word, one subject and one object. The subject is regarded as a potential source and the object is considered as a potential quotation. To ensure that the quotations are informative, we also require that the length of the object should be more than three words.

\paragraph{Source and Quote Filtering}

We required that the subject of a candidate sentence should be a person or an organisation and therefore identified potential source entities via the accompanying DBpedia ontology labels\footnote{\url{http://mappings.dbpedia.org/server/ontology/classes/}} in the dataset. Our selected ontology classes are shown in our source code repository \footnote{\url{https://github.com/WenjiaZh/NewsQuote/blob/main/SelectedOntologyClasses.txt}}.
Since each entity could have more than one ontology class, we further removed samples with sources labeled as \emph{Location}, \emph{Place} and \emph{Country}. As the same subject could have multiple mentions, we use DBPedia entity links for entity resolution and normalisation. In addition, 
we required a named entity to appear at least twice in the dataset. Finally, to avoid the sentence split error, we required quotation marks to be paired in sentences that contain direct quotes and mixed quotes. 

\begin{figure*}[ht]
    \centering
    \includegraphics[width=0.9\textwidth]{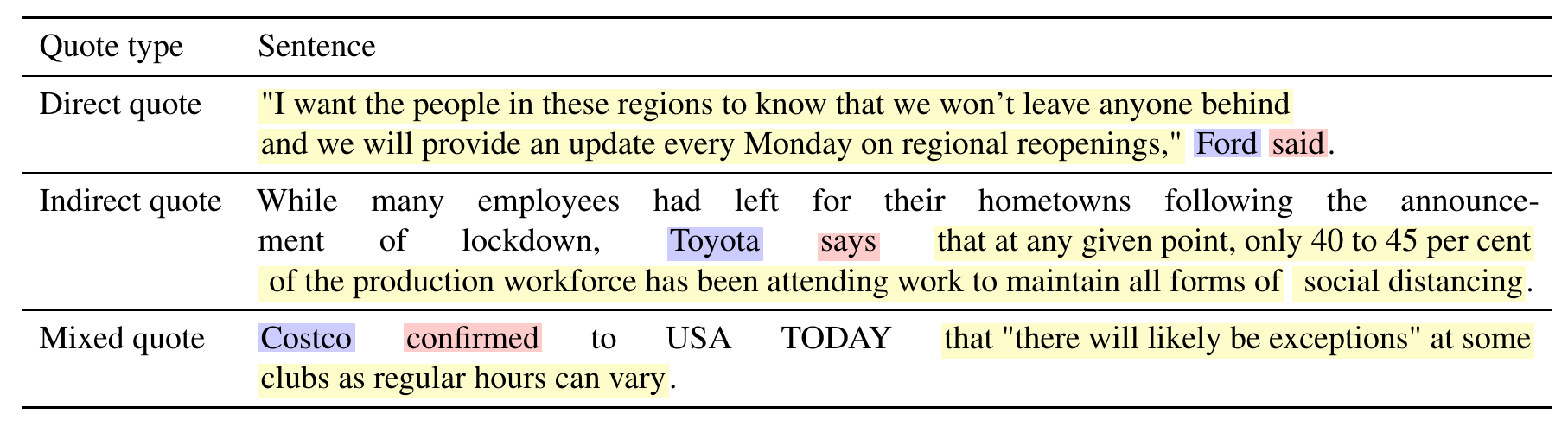} 
    \caption{Three types of quotes in our dataset. Sources are highlighted in blue, trigger verbs are highlighted in red, and quotes are highlighted in yellow.}
    \label{tab:testsamples}
\end{figure*}

\subsection{Test Set Annotation}
\begin{figure} [ht]
\centering
\includegraphics[width=0.47\textwidth]{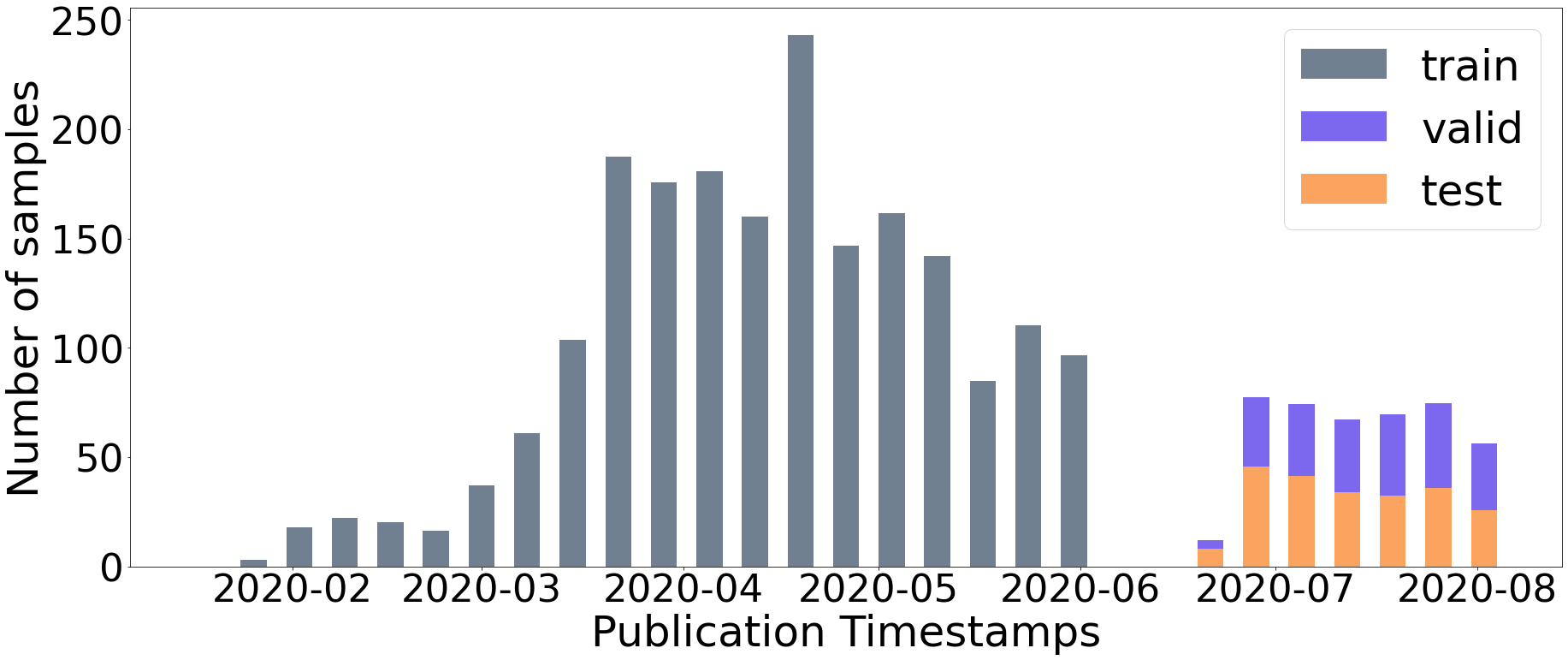} 
\caption{The distribution of quote-source pairs. The training set contains samples released from 2020-01-19 to 2020-05-31, and the validation/testing set contains samples released from 2020-06-21 to 2020-08-02.} 
\label{Fig.time} 
\end{figure}

Since in practice, given a topic, we can only identify experts based on their previous quotes published in earlier news articles, 
we divide the dataset into training, validation and testing sets by 
news articles publishing timestamps, ensuring quote-source pairs in the validation and testing sets occurred later than those in the training set. 
Figure \ref{Fig.time} demonstrates the distribution of quote-source pairs based on the publishing dates of their associated news articles.\footnote{There is no data between 2020-05-31 and 2020-06-21 in the original dataset.} 

To ensure data quality, samples in the test set were manually screened 
by one annotator. We list five types of noise and corresponding examples appearing in the raw test set in Table \ref{tab:noisy}. Data falling into one of these noise categories were removed from the test set. 

\subsection{Dataset Statistics}
Our data covers three categories of quotes, illustrated in Figure \ref{tab:testsamples}. In short, direct quotations are placed inside quotation marks, while indirect quotations are not, and a mix of direct and indirect quotations have only part of the quotations placed inside quotation marks. We roughly estimated the weight of each quotation type on the dataset by the number and position of quotation marks: 81\% for indirect quotes, 10\% for direct quotes and 9\% for mixed quotes. 
In the test set, there are 1,867 (84\%) indirect quotes, 215 (10\%) mixed quotes and 143 direct quotes (6\%).  
Table \ref{tab:datasta} shows the statistics of our final \textsc{NewsQuote} dataset. In summary, we have a total of 24,031 English source-quote pairs with 3,246 sources from 258 global sources. More related statistics and plots are presented in Appendix \ref{sec:appendix-DataStat}.

\begin{table}[h]
\centering
\begin{tabular}{l c c c}
\toprule
 &  \textbf{Test} & \textbf{Valid} & \textbf{Train} \\ 
\midrule
\textbf{No. of samples} & 2,236 & 2,082 & 19,713 \\
\textbf{No. of articles} & 1,937 & 1,766 & 14,526 \\
\textbf{No. of source entities} &  1,016 & 765 & 2,963 \\
\textbf{Avg. quote length} & 28.38 & 29.16 & 28.99 \\
\textbf{No. of news sources} & 180 & 178 & 252 \\
\textbf{No. of news categories} & 470 & 440 & 629 \\
\textbf{Avg. keywords per article} &  43.23 & 44.28 & 42.17\\
\bottomrule
\end{tabular}
\caption{The \textsc{NewsQuote} Dataset statistics.}
\label{tab:datasta}
\end{table}

\begin{figure*}[ht]
    \centering
    \includegraphics[width=0.9\textwidth]{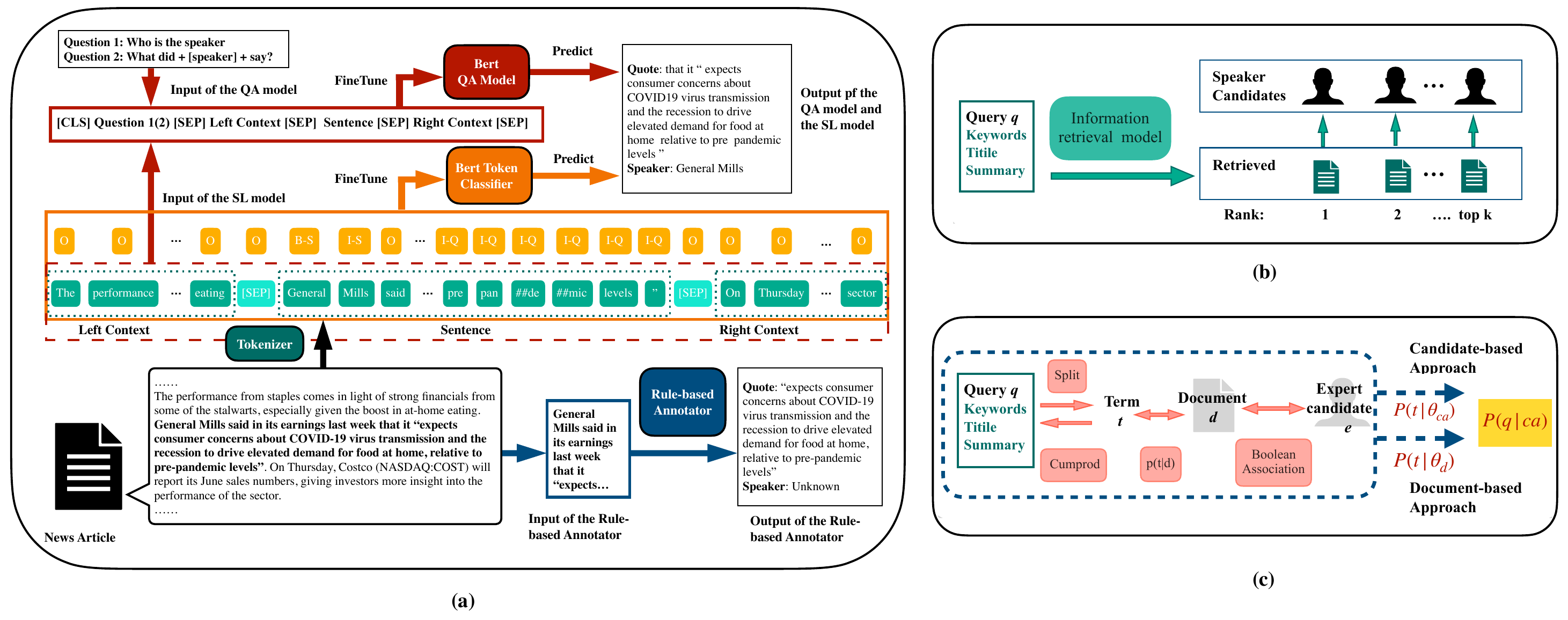} 
    \caption{Illustrations of 5 approaches described in Section \ref{sec:BA}. Plot(a) describes the QA pipeline, the sequence labelling and the Rule-based Quota Annotator used for quote-source extraction. Plot(b) introduces the document retrieval approach for expert recommendation, and plot(c) presents the expert retrieval approach for expert recommendation}
    \label{fig:approach}
\end{figure*}

\section{Task Definition}
In our dataset, each sample $S_{i}$ consists of a context  $c_{i}$, a quote-source pair $(q,e)_{i}$, a list of keywords $k_{i}$ and metadata $m_{i}$. The context contains 3 sentences, the main sentence where the source and quote appeared, its preceding sentence and following sentence. Both keywords and metadata are defined at the document level and are retrieved from the AYLIEN coronavirus dataset. 
We propose the following two tasks on this \textsc{NewsQuote} dataset:

\noindent\textbf{Source and quote extraction} 
is defined as automatically extracting the source-quote pair $(q,e)_{i}$ from a given context $c_{i}$. 

\noindent\textbf{Expert recommendation} involves suggesting a ranked list of experts given a query, based on what they said in the past. 

\section{Approaches}\label{sec:BA}
We present approaches for source and quote extraction, and expert recommendation. An overview of the approaches is illustrated in Figure \ref{fig:approach}. 

\subsection{Source and Quote Extraction}
We tackle the problem of extracting quote-source pairs using three approaches: rule-based method, sequence labelling, and question answering.

\paragraph{Approach 1: Rule-based Quote Annotator}
Regular-expression-like rules can be used to extract direct quotes. We  run the Quote Annotator \footnote{\url{https://stanfordnlp.github.io/CoreNLP/quote.html}} from Stanford CoreNLP \cite{manning-etal-2014-stanford}  on our test sample sentences. It can only extract direct quotes that are delimited by quotation marks.

\paragraph{Approach 2: Sequence Labelling}
We can label each sample in our dataset with a 5-class BIO tagging scheme. The source is annotated by 'B-S' and 'I-S', denoting whether the corresponding token indicates the beginning of a source mention, or is inside a source mention. Similarly, the quotation is annotated by 'B-Q' and 'I-Q', and all the other tokens are marked by 'O'. We then fine tune a BERT-based token classifier \cite{devlin2018bert} to identify sources and quotes from the context. 

\paragraph{Approach 3: Question Answering (QA) pipeline}
\label{QAmethod}
We use a QA pipeline for source and quote extraction by asking two questions in turn: 
\begin{quote}
Q1: Who is the source?\\
Q2: What did [source] say?
\end{quote}
During training, the \texttt{[source]} in \texttt{Q2} is the gold standard answer for question \texttt{Q1}. During inference, it is the extracted answer for \texttt{Q1}. The input context is composed of a question, a left sentence, $l$, a main sentence, $s$ and a right sentence, $r$. To extract the answer from the context, we fine-tuned the pre-trained BERT-based extractive QA model \cite{devlin2018bert}, where the input takes the form:
\begin{quote}
[CLS] Question [SEP] l [SEP] s [SEP] r [SEP]
\end{quote}

\subsection{Expert Recommendation}
\label{ERmethods}

We can formulate expert recommendation as a retrieval problem, that given a query, we would like to retrieve sources who can comment on the topic discussed in the query ranked by their relevance to the query. There are two possible approaches, one is to use sources' past quotes as documents and perform \emph{document retrieval} and then return the sources of the retrieved quotes as results, another is to perform \emph{expert retrieval} directly.

\paragraph{Approach 1: Document Retrieval}
\emph{Document retrieval} aims to first retrieve relevant documents (i.e., the context where a quote appears) given a query, and then extract the sources from the documents as results. 
For document indexing, we experiment with a sparse bag-of-words Lucene index and four kinds of dense transformer-encoded Faiss indices via Pyserini\footnote{\url{https://github.com/castorini/pyserini}}. A BM25 ranking approach on the sparse index and a nearest-neighbor search on dense indexes were then applied to return the top 10 most relevant documents for a given query. Sources in the top 10 retrieved documents are then identified as the recommended experts. 

\paragraph{Approach 2: Expert Retrieval}

\emph{Expert retrieval} directly retrieves sources based on the 
probability of a query conditional on a given candidate source $P(q|e)$. Following the the framework introduced by \citet{balog2009language}, we implemented both candidate-based and document-based expert finding approaches. 

\noindent\underline{Candidate-Based Expert Retrieval} Assuming that each term in the query is sampled identically and independently, also that the document and the expert source candidate are conditionally independent, the candidate-based approach estimates $P(q|e)$ by: 

\begin{gather}
P(q|e)=\prod_{t\in q} \{ (1-\lambda)(\sum_{d\in D}p(t|d)p(d|e)+\lambda p(t) \} ^{n(t,q)},\nonumber \\
\lambda=\frac{\beta}{\beta+n(e)},  \quad
\beta=\frac{\sum_{E}|\{d: n(e,d)>0\}|\cdot|d|}{|E|}, \nonumber
\end{gather}
\noindent where $\lambda$ is the smoothing parameter, $p(t|d)$, $p(d|e)$ and $p(t)$ are the conditional probability of a term $t$ in document $d$, the conditional probability of a document $d$ given source $e$, and the probability of term $t$, respectively. Both $p(t|d)$ and $p(t)$ are estimated by maximum likelihood. The probability $p(d|e)$ is set by a Boolean model, which will be discussed later. $|d|$ is the average document length, $n(t,q)$ is the number of times that a term $t$ appears in the query $q$, $n(e,d)$ is the occurrence frequency of an expert $e$ appeared in the document $d$, and $n(e)$ is the total number of occurrences in documents associated with the source $e$.

\noindent\underline{Document-Based Expert Retrieval} 
The document-based expert retrieval approach searches for sources via relevant document collection. This approach assumes the conditional independence between the query and candidate, and estimates the probability of a term $t$ in each document: 

\begin{small}
\begin{gather}
\small
  P(q|e)=\sum_{d\in D} \{\prod_{t\in q}((1-\lambda)p(t|d)+\lambda p(t))^{n(t,q)}\}p(d|e), \nonumber \\
\lambda=\frac{\beta}{\beta+n(d)},  \quad
\beta=|d|, \nonumber
\end{gather}
\end{small}
\noindent where $n(d)$ is the length of document $d$.

In both the candidate-based and document-based expert finding approaches, 
the document-candidate associations, $p(d|e)$, is estimated by a simple Boolean model, where it is set to 1, if $n(e,d)>0$, and 0, otherwise. 

\section{Experiments}

\subsection{Experimental Setup}

For the rule-based approach, we directly feed the raw sentences into the Quote Annotator. To build the token classifier, we segment the input text into a sequence of 512 tokens, and fine tune the model for 100 epochs with an initial learning rate of 2e-7. For the extractive QA model, the maximum length of the extracted answer is set to 30 when questioning sources and 512 when questioning quotes. For the question about source, we train the model for 50 epochs with an initial learning rate of 2e-6. For the question about quote, we train the model for 100 epochs with an initial learning rate of 2e-5. 

For expert recommendation, we consider two types of documents: the main sentence where a source/quote occurred, or the main sentence together with its surrounding context (i.e., the preceding and following sentences). 
For the query to be used for expert retrieval, we use either the title of a news article, its keywords , or the first sentence of the summary. To further remove interference, we eliminate the source name from the input query if there is any. 
For the expert retrieval method, we take only the first $w$ words in the news article title (the keyword list or the first sentence of the news summary) as the input query to reduce the running time. After validating the value of $w$ between 1 and 10, we finally set $w = 5$. 

\begin{table*}[htb]
\centering
\resizebox{0.9\linewidth}{!}{
\setlength{\tabcolsep}{1mm}{
\begin{tabular}{lcccccccc}
\toprule
 &\multicolumn{2}{c}{\textbf{Overall}} &\multicolumn{2}{c}{\textbf{Direct Quotes}}&\multicolumn{2}{c}{\textbf{Indirect Quotes}} &\multicolumn{2}{c}{\textbf{Mixed Quotes}}\\
\cmidrule(lr){2-3}\cmidrule(lr){4-5}\cmidrule(lr){6-7}\cmidrule(lr){8-9}
 & \textbf{Macro F1} & \textbf{Exact Match}& \textbf{Macro F1} & \textbf{Exact Match}& \textbf{Macro F1} & \textbf{Exact Match}& \textbf{Macro F1} & \textbf{Exact Match}\\ 
 \midrule
 \textbf{Rule$_{source}$} & 5.76 & 5.62  & 50.58 & 49.65 & 0.214 & 0.214 &24.11&23.26 \\
 \textbf{Rule$_{quote}$} & 7.72 & 1.93 & 82.33 &30.07&0.145&0.00&23.84&0.00 \\
\midrule
\textbf{SL$_{source}$} &98.06&95.37 & 98.63& 95.80& 97.99& 95.34& 98.23& 95.35 \\
\textbf{SL$_{quote}$} & 95.65&85.17&\textbf{97.17}&89.51&95.61&85.11&95.05&82.79 \\
\midrule
\textbf{QA$_{speakr}$} & \textbf{98.86} & \textbf{98.61}&\textbf{99.30}&\textbf{99.30} & \textbf{98.77}&\textbf{98.50}&\textbf{99.38}&\textbf{99.07}\\ 
\textbf{QA$_{quote}$} &  & \\ 
\textbf{$~~_{w/~true~source}$} & \textbf{95.96} & \textbf{90.74}&95.83& \textbf{93.01}&\textbf{95.96}&\textbf{90.31}&\textbf{96.06}&\textbf{93.02}\\ 
\textbf{$~~_{w/~pred.~source}$} & 95.61 & 89.93 & 95.78&93.01&95.55&89.34&96.06&93.02\\ 
\textbf{$~~_{w/~source~mask}$} & 93.92 & 85.84&96.53&90.21&93.56&85.11&95.28&89.30\\ 
\toprule
\end{tabular}}}
\caption{Results of source and quotation extraction on the test set. \textbf{Rule} $-$ the rule-based annotator, \textbf{SL} $-$ sequence labeling, \textbf{QA} $-$ the question answering pipeline. The subscripts indicate the aim of the models, either for ${source}$ extraction or for ${quote}$ extraction. Under the QA$_{quote}$, `$_{w/~true~source}$' is where we use the true source name when asking "\emph{What did + [source] + say?}", while `$_{w/~pred.~source}$' uses the predicted source from the QA$_{speakr}$ results, and `$_{w/~source~mask}$' uses the generic word "\emph{they}".}
\label{tab:QAresults}
\end{table*}

\subsection{Evaluation Metrics}

To measure model performances for quote extraction and attribution, we use two metrics defined in SQuAD \cite{rajpurkar-etal-2016-squad}, the exact match and the macro-averaged F1 score. \textbf{Exact Match} is equal to one if the predicted outcome is completely identical to the ground truth, while 
\textbf{(Macro-averaged) F1} measures the average overlap between predicted and ground truth answers at the token-level. 

For expert recommendation, we use two metrics commonly used in information retrieval, the mean average precision (MAP) and the normalized discounted cumulative gain (NDCG). \textbf{Mean Averaged Precision} is the average value of the precision at the points where relevant documents are retrieved. 
\textbf{Normalized Discounted Cumulative Gain at K} first discounts the gain scale at the $i$-th rank position by $\frac{1}{\log_{2}(i)}$, then adds up the converted gain scales up to rank $k$, and finally normalizes the result by the ideal ranking order. 
In addition, we propose \textbf{relaxed metrics} where the retrieved expert is considered relevant if it is in the same cluster as the true source. In the construction of relaxed metrics, we opt for the top 100 most frequent source DBpedia categories and use the binary vectors to embed sources\footnote{In our dataset, a source is assigned to 4 to 5 DBpedia categories on average.}. We then perform $k$-means clustering on the source embeddings. We empirically set $k=40$ according to the cluster coherence and separation scores. 


\begin{table*}
\small
\centering
\resizebox{0.6\linewidth}{!}{
\begin{tabular}{l|ccc|ccc}
\toprule
 & \multicolumn{3}{|c|}{\textbf{Strict Metrics}}& \multicolumn{3}{|c}{\textbf{Relaxed Metrics}} \\
& \textbf{MAP} & \textbf{NDCG$_{5}$} & \textbf{NDCG$_{10}$}& \textbf{MAP} & \textbf{NDCG$_{5}$} & \textbf{NDCG$_{10}$}\\ 
\midrule
\textbf{DR$_{sparse}$} & \textbf{0.2903} & \textbf{0.2807} & \textbf{0.3590} &  \textbf{0.4162} &  \textbf{0.3925} &  \textbf{0.5183} \\ 
\textbf{DR$_{flat}$} & 0.1481 & 0.1440 & 0.1939 & 0.2886 & 0.2714 & 0.3887 \\ 
\textbf{DR$_{hnswpq}$} & 0.1509 & 0.1473 & 0.1926 &0.2966 & 0.2805 & 0.3956\\ 
\textbf{DR$_{hnsw}$} & 0.1446 & 0.1406 & 0.1889 & 0.2865 &0.2686 & 0.3850\\ 
\textbf{DR$_{pq}$} & 0.1395 & 0.1363 & 0.1838 &0.2739 &0.2583 & 0.3734\\ 
\midrule
\textbf{ER$_{can}$} & 0.1021 & 0.1106 & 0.1252 & 0.2306 & 0.2294 & 0.3135\\ 
\textbf{ER$_{doc}$} & 0.1205 & 0.1281 & 0.1418 & 0.2465 & 0.2412 & 0.3285\\ 
\toprule
\end{tabular}}
\caption{Results of expert recommendation using quote context as document, and news article keywords as query. In the first five rows, \textbf{DR} denotes the document retrieval approach, and the subscripts represent 5 types of retrieval indices mentioned in Section \ref{ERmethods} Approach 1, Lucene sparse bag-of-words index, Faiss flat index, Faiss HNSWPQ index, Faiss HNSW index, and Faiss PQ index. \textbf{ER}$_{can}$ is the candidate-based expert finding approach, and \textbf{ER}$_{doc}$ is the document-based expert finding approach. In document-based expert finding approaches, the input query length is set to 5 keywords.} 
\label{tab:expertRec}
\end{table*}

\subsection{Experimental Results}
We first present the results of the three quote extraction and attribution methods described in Section \ref{QAmethod}, and subsequently present the evaluation results for the two expert recommendation approaches introduced in Section \ref{ERmethods}.

\paragraph{Quote Extraction and Attribution}

Table \ref{tab:QAresults} presents the performance of the rule-based annotator, sequence labeling and the QA pipeline 
on the test set. It is not surprising that the rule-based quote annotator performs the worst as it can only extract direct quotes using regular-expression-like rules. In our test set, only 337 out of 2225 samples were identified as containing quotes. On this subset, the rule-based annotator gives a higher exact match score of 49.65 for sources compared to quotes. But it performs much better for direct quote extraction in Macro F1 compared to source extraction. On the other two categories, indirect and mixed quotes, the rule-based annotator essentially failed to produce any sensible results. 
Sequence labeling gives much better results compared to the rule-based annotator. We notice that in terms of exact match, quote extraction appears to be nearly 10\% lower than source extraction, showing that the model struggled with longer answer extraction. For the three categories of quotes, the model gives the best results for quote extraction on the direct quotes, followed by the indirect quotes, and it performs the worst on the mixed quotes. This is expected since mixed quotes are more complex to deal with compared to the other two categories. 
The QA pipeline achieves the best performance in both identifying sources and extracting the quotations. 
In testing the QA pipeline's quote extraction capabilities, we experimented with three scenarios by using either: the true source name in the question for quote, the predicted source from the results of QA$_{source}$, 
or masking the source with the pronoun '\emph{they}' to completely remove the source information from the question.  
Since the accuracy of our QA model for source identification is already high, using the true or predicted source for the question for quote extraction does not make much difference. However, if the source information is lost, the quote extraction performance drops by nearly 2\% in Macro F1 and over 4\% in exact match. 

\paragraph{Expert Recommendation}


We show in Table \ref{tab:expertRec} the expert recommendation results from using keywords of a news article as query, and the context of quotes (the main sentence where the source and quote occurred, together with the preceding and the following sentences) as the document. It can be observed that the document retrieval (\textbf{DR}) approaches generally outperform the expert retrieval (\textbf{ER}) approaches. Among various document indexing strategies, using Lucene sparse bag-of-words index (\textbf{DR$_{sparse}$}) gives superior results compared to other dense transformer-encoded Faiss indices. As expected, using the Relaxed Metrics where a retrieved source is considered as relevant to the true source if they reside in the same cluster, we obtain better results compared to the strict metrics.\footnote{Results using other document retrieval or expert retrieval approaches based on different combinations of the formulation of documents and queries are in Appendix \ref{sec:appendix-results}.}

\section{Challenges and Future Directions}

We have presented our NewsQuote dataset, and introduced a new task, namely expert recommendation in the field of journalism and fact-checking. Our experiments confirmed the possibility of extracting quote-source pairs using a question-answering pipeline as well as finding expert sources using document retrieval and expert retrieval. Here, we outline some potential future directions. 

First, in the construction of our dataset, the quote trigger verbs are manually selected from the most frequent group of verbs. On one hand, the identified verb list does not cover all the possible verbs that are indicative of quotations, such as those occurred less frequently or are not closely related to the Covid topic. On the other hand, some verbs are ambiguous and need to be contextualized to determine whether they are indeed the trigger words. Although we removed disambiguous cases when examining the test set, it is not practical to perform manual filtering on such large-scale data. Future work could explore the possibility of leveraging other large-scale quote corpora for training a model for the detection of quote trigger words. Also, our dataset has been constructed from the news articles about the coronavirus. In the future, this could be extended to cover a wide range of topics such as business, technology, education, and politics. 

Second, co-reference resolution will be vital for increasing the quote-source attribution data as it is common to use pronouns to refer to previously mentioned sources in news articles. Our preliminary experiments on co-referencing resolution led to noisy quote-source attribution results. In the future work, the content similarity and/or coherence between the quote associated with a pronoun and a quote of a candidate source could be leveraged to improve the co-reference resolution results. 

Third, with the DBpedia links referred to as identifications of sources in our dataset, external knowledge could be imported as evidence to enhance the performance of expert recommendation. 

Fourth, our framework makes it possible to build a quote-source library for the newsroom that can help with veracity assessment, where summaries of the comments made by each source, including who has quoted them, when and in relation to which veracity check, can be made available to journalists and fact-checkers, thereby reducing duplication of effort and supporting collaboration.

Finally, it is important that journalists and fact-checkers do not become over-reliant on tools such as the one we present here (i.e., fall victim to so-called 'automation bias'). The results therefore need to be interpreted with care and the final decision on which experts to approach should always made by the journalist or fact-checker. It is therefore important that such models provide evidence for their recommendations that can be assessed for credibility and relevance by the user \cite{procter2023}.


\section{Conclusions}

We have described the construction of a novel, large-scale dataset on quote-source pairs retrieved from news articles. 
Our \textsc{NewsQuote} dataset comprises direct quotations, indirect quotations and their combinations. The diversity of quote types will encourage the development of more advanced approaches for the challenging tasks of indirect and mixed quote extraction. 
Based on the \textsc{NewsQuote} dataset, we have demonstrated that the QA pipeline is able to achieve over 98\% exact match for source extraction and close to 90\% for quote extraction. In addition, we have introduced the expert recommendation task and shown that the document retrieval approach with sparse indexing gives the best results compared to other dense retrieval approaches.  

\section*{Ethics Statement}
All data we used are from open public sources. We have obtained a written consent from the Aylien to download their data. As per the data owner's requirement, we will not directly share the downloaded data, instead, we will share the download script and all pre-processing scripts so that others could obtain the same dataset we used in the paper from the Aylien's website. 


\section*{Acknowledgements}
This work was supported in part by the EPSRC (grant no. EP/V048597/1). YH is supported by a Turing AI Fellowship funded by the UKRI (grant no. EP/V020579/2).

\bibliography{bibliography}

\appendix

\setcounter{table}{0}
\renewcommand{\thetable}{A\arabic{table}}
\setcounter{figure}{0}
\renewcommand{\thefigure}{A\arabic{figure}}

\begin{figure*}[t]
    \centering
     \begin{subfigure}[t]{0.3\textwidth}
         \centering
         \includegraphics[width=\textwidth, height=4.5cm]{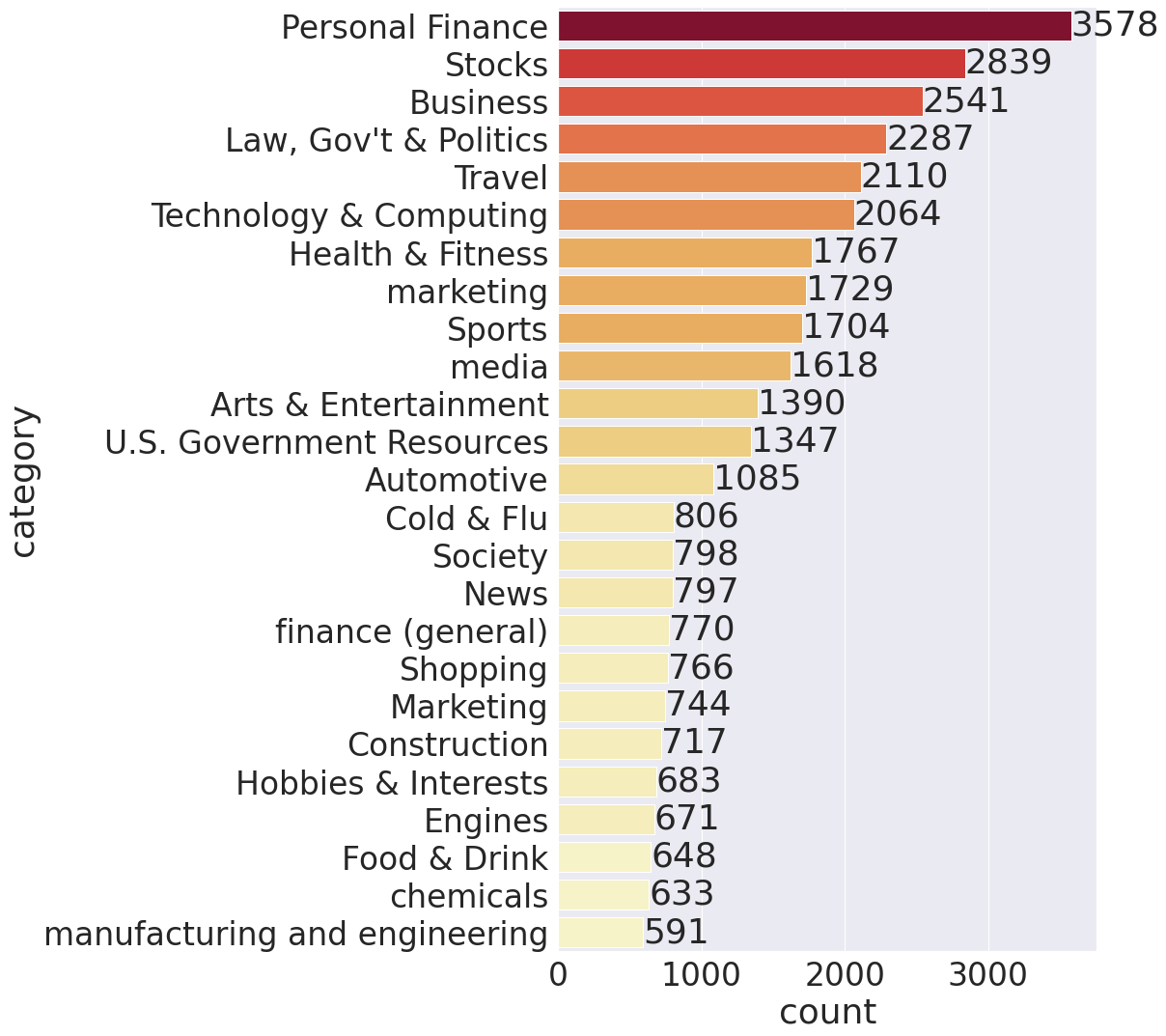}
         \caption{Top frequent news article categories. The count is the number of articles in the corresponding category.}
         \label{AC1}
     \end{subfigure}
     \hfill
     \begin{subfigure}[t]{0.3\textwidth}
         \centering
         \includegraphics[width=\textwidth, height=4.5cm]{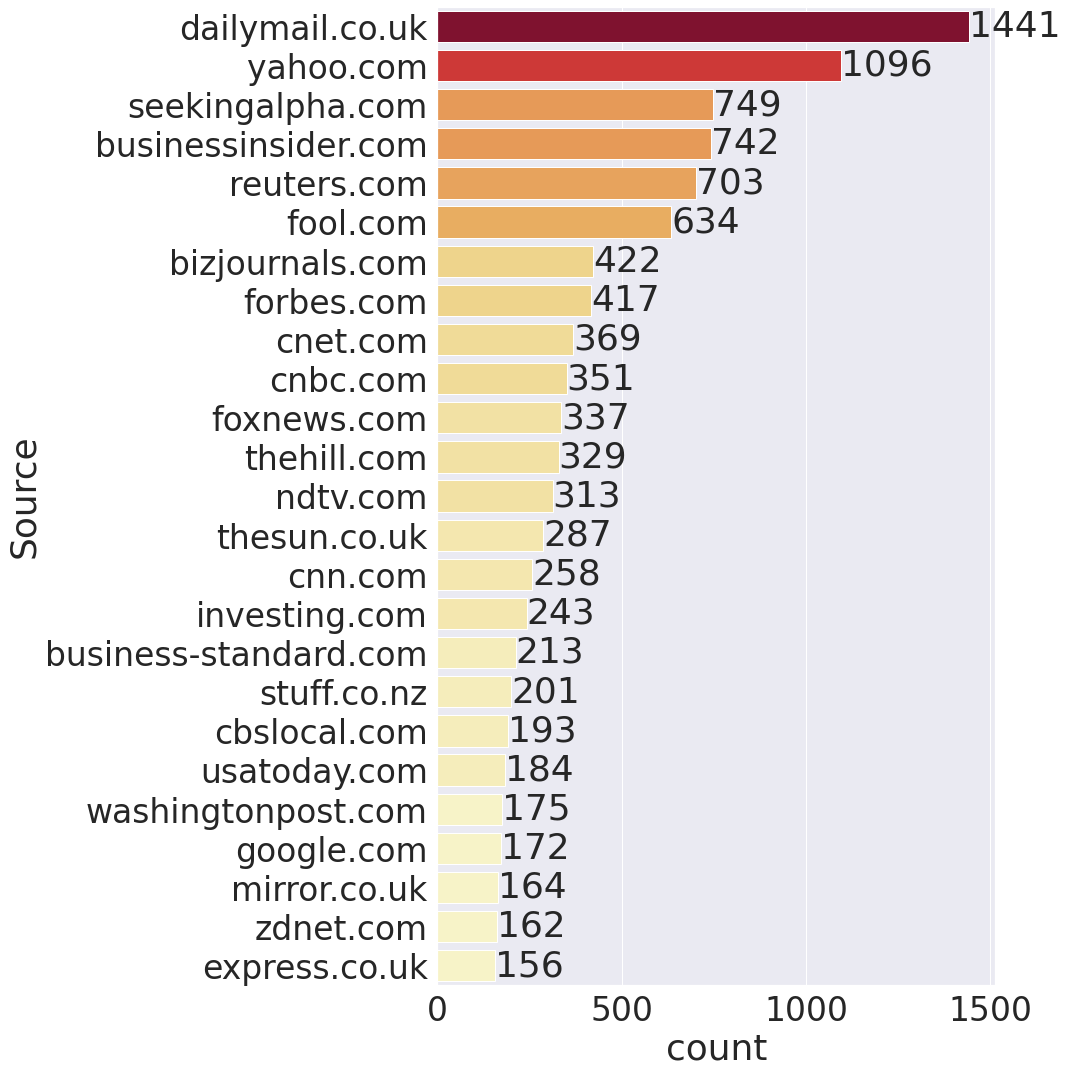}
         \caption{Top news sources. The count is the number of articles published by the corresponding news source.}
         \label{AC2}
     \end{subfigure}
     \hfill
     \begin{subfigure}[t]{0.3\textwidth}
         \centering
         \includegraphics[width=\textwidth, height=4.5cm]{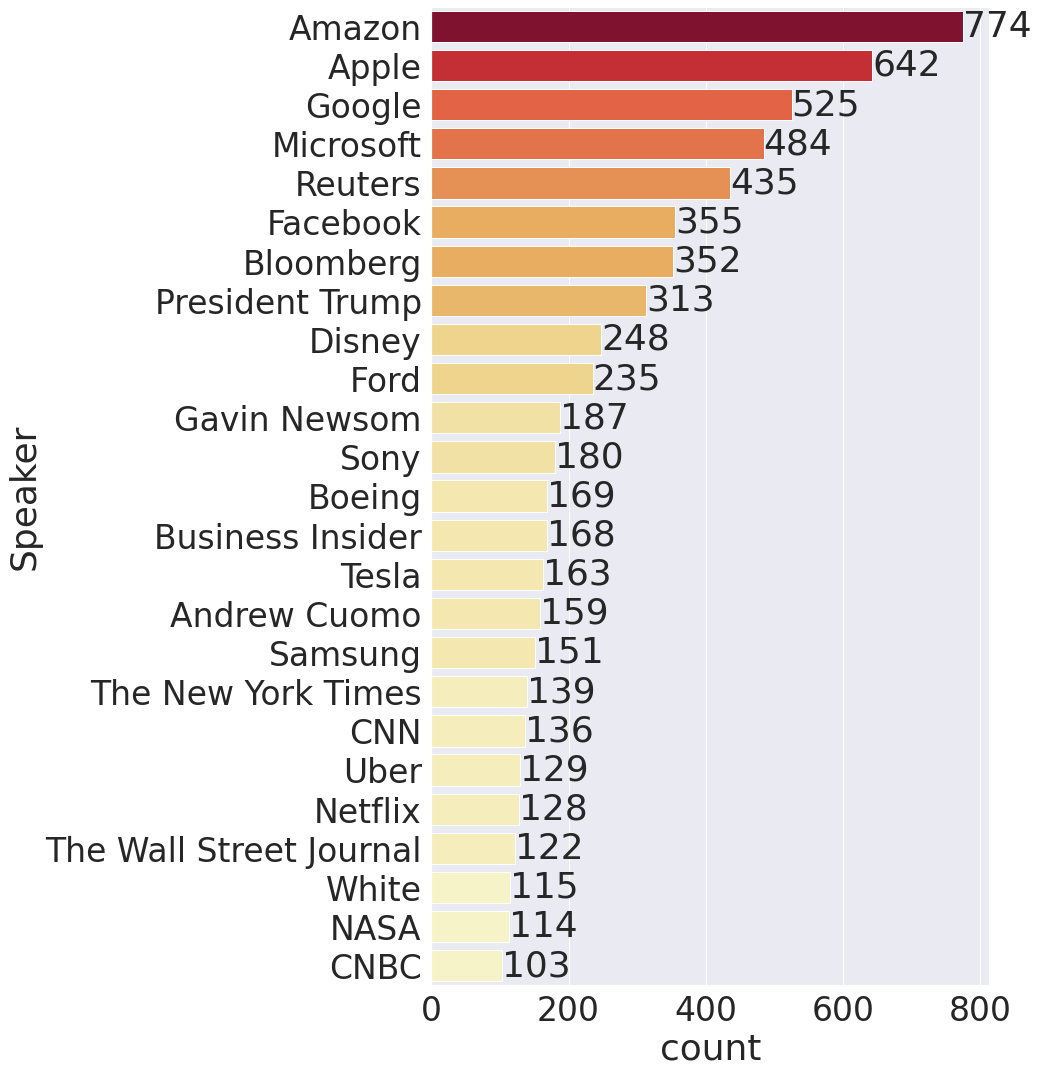}
         \caption{Top frequent sources in our quote-source pair dataset. The count is the number of quotations came from the corresponding sources.}
         \label{AC3}
     \end{subfigure}
     \caption{ }
        \label{fig:data statistics}
\end{figure*}

\begin{table*}[!h]
\small
\centering
\begin{tabular}{p{3.8cm} p{12.8cm}}
\toprule
\textbf{Noise Type} & \textbf{Example Text} \\
\midrule
Jumble text &  -=-=-=- +++lead-in-text Last August, Apple announced that it would [distribute special iPhones](https://www.\\
\midrule
Incorrect labeling of the quote (only text in bold is marked as quote) & Furthermore, CBS reported, citing \textbf{current and former diplomats} with insight into the situation, that Gunter since his nomination in May 2019 has created an increasingly "untenable" working environment by "flying into a rage" and changing deputy chiefs of mission at will.\\
\midrule
Improper trigger verb & "These are warnings that have been inevitable from the very start and exactly the reason why ICE should have, and should continue to, \textbf{release} people, especially those who are medically vulnerable to COVID-19, to prevent a humanitarian disaster," she said.\\
\midrule
Improper source &  We are in professional corporate relations with various companies and this helps us in digging out \textbf{market data} that helps us generate accurate research data tables and \textbf{confirms utmost accuracy in our market forecasting}.\\
\midrule
 Not an affirmative statement & Apple \textbf{didn't respond} to a request for comment.\\
\bottomrule
\end{tabular}
\caption{Types of noisy samples removed from the test set.}
\label{tab:noisy}
\end{table*}

\section{Data Statistics}
\label{sec:appendix-DataStat}


Figure \ref{AC1} presents 25 of the most frequent news article categories. One average, each article has 3.92 category labels. The top five categories are: 'Personal Finance', 'Stocks', 'Business', 'Law, Gov't \& Politics', and 'Travel'.


Figure \ref{AC2} shows 25 of the most common news sources. In our dataset, the total number of source platforms is 258. The top 5 source platforms are: Daily Mail, Yahoo, Seeking Alpha, Business Insider, Reuters.


Figure \ref{AC3} lists 25 of the most frequent sources. in total we have 3246 sources. The top 5 sources are: Amazon, Apple, Google, Microsoft, Reuters.

\begin{table*}[!h]
\small
\centering
\setlength{\tabcolsep}{1.5mm}
\resizebox{1.5\columnwidth}{!}{
\begin{tabular}{lllcccccc}
\toprule
 & & & \multicolumn{3}{c}{\textbf{Strict Metrics}}& \multicolumn{3}{c}{\textbf{Relaxed Metrics}} \\
 \textbf{Document}& \textbf{Query}& \textbf{Method} & \textbf{MAP} & \textbf{NDCG$_{5}$} & \textbf{NDCG$_{10}$}& \textbf{MAP} & \textbf{NDCG$_{5}$} & \textbf{NDCG$_{10}$}\\
\midrule
& & \textbf{DR$_{sparse}$} & \textbf{0.2903} & \textbf{0.2807} & \textbf{0.3590} &  \textbf{0.4162} &  \textbf{0.3925} &  \textbf{0.5183} \\ 
& & \textbf{DR$_{flat}$} & 0.1481 & 0.1440 & 0.1939 & 0.2886 & 0.2714 & 0.3887 \\ 
& & \textbf{DR$_{hnswpq}$} & 0.1509 & 0.1473 & 0.1926 &0.2966 & 0.2805 & 0.3956\\ 
 \textbf{Context} &  \textbf{Keyword} &\textbf{DR$_{hnsw}$} & 0.1446 & 0.1406 & 0.1889 & 0.2865 &0.2686 & 0.3850\\ 
& & \textbf{DR$_{pq}$} & 0.1395 & 0.1363 & 0.1838 &0.2739 &0.2583 & 0.3734\\ 
& & \textbf{ER$_{can}$} & 0.1021 & 0.1106 & 0.1252 & 0.2306 & 0.2294 & 0.3135\\ 
& & \textbf{ER$_{doc}$} & 0.1205 & 0.1281 & 0.1418 & 0.2465 & 0.2412 & 0.3285\\ 
\midrule
& & \textbf{DR$_{sparse}$} & 0.1357 & 0.1367 & 0.1719 & 0.3112& 0.2980&0.4048 \\ 
& &\textbf{DR$_{flat}$} & 0.1321 & 0.1304 &  0.1670  & 0.2979& 0.2800&0.3917\\ 
& &\textbf{DR$_{hnswpq}$} & 0.1206 & 0.1177 & 0.1538  & 0.2830 & 0.2659 &0.3790\\
 \textbf{Context} &  \textbf{Title} &\textbf{DR$_{hnsw}$} & 0.1311 & 0.1299 & 0.1659  & 0.2963 & 0.2787&0.3902\\ 
& &\textbf{DR$_{pq}$} & 0.1231 & 0.1223 & 0.1570 & 0.2931 &0.2797 &0.3873\\ 
& &\textbf{ER$_{can}$}& 0.1062&0.1116 & 0.1289  & 0.2516& 0.2485&0.3366\\ 
& &\textbf{ER$_{doc}$} &0.1111 &0.1183 &0.1320  & 0.2532& 0.2518&0.3387\\ 
\midrule
& &\textbf{DR$_{sparse}$} & 0.1498 & 0.1484 & 0.1884   & 0.3285&0.3115 &0.4297\\ 
& &\textbf{DR$_{flat}$} & 0.1435 & 0.1389 &  0.1818   & 0.3125&0.2945 &0.4089\\ 
& &\textbf{DR$_{hnswpq}$} & 0.1285 & 0.1254 & 0.1648   & 0.2969& 0.2761&0.3953\\ 
\textbf{Context} &  \textbf{Summary} &\textbf{DR$_{hnsw}$} & 0.1419 & 0.1373 & 0.1803  & 0.3097& 0.2907&0.4056\\ 
& &\textbf{DR$_{pq}$}  & 0.1402 &0.1418 &0.1773  & 0.3103& 0.2953&0.4098\\ 
& &\textbf{ER$_{can}$}  & 0.0918&0.0966 & 0.1157 & 0.2461&0.2419 &0.3337\\ 
& &\textbf{ER$_{doc}$}  & 0.1014&0.1069 &0.1235  & 0.2570& 0.2542&0.3479\\ 

\midrule

& &\textbf{DR$_{sparse}$} & 0.1356 & 0.1366 & 0.1730  & 0.3059& 0.2938&0.4074\\ 
& &\textbf{DR$_{flat}$} & 0.0721 & 0.0724 &  0.0999  & 0.2279& 0.2127&0.3238\\ 
& &\textbf{DR$_{hnswpq}$} & 0.0756 & 0.0746 & 0.1027  & 0.2501& 0.2410 & 0.3458\\ 
\textbf{Quote} &  \textbf{Keyword} &\textbf{DR$_{hnsw}$} & 0.0713 & 0.0719 & 0.0984  & 0.2266& 0.2124&0.3214\\ 
& &\textbf{DR$_{pq}$} & 0.0692 & 0.0684 & 0.0931 & 0.2161 &0.2036 &0.3047\\ 
& &\textbf{ER$_{can}$} & 0.0685 & 0.0720 & 0.0815 & 0.1902& 0.1902&0.2633\\ 
& &\textbf{ER$_{doc}$} & 0.0552& 0.0588& 0.0682 & 0.1685& 0.1647& 0.2404\\ 

\midrule
& &\textbf{DR$_{sparse}$} & 0.0955 & 0.0956  & 0.1204  & 0.2713& 0.2614&0.3600\\ 
& &\textbf{DR$_{flat}$} & 0.0953 & 0.0958 & 0.1227  & 0.2744& 0.2611&0.3729 \\ 
& &\textbf{DR$_{hnswpq}$} & 0.0861 & 0.0865 & 0.1113  & 0.2683& 0.2526&0.3654\\ 
\textbf{Quote} &  \textbf{Title}&\textbf{DR$_{hnsw}$} & 0.0943 & 0.0946 & 0.1209  & 0.2751& 0.2620&0.3724\\ 
& &\textbf{DR$_{pq}$} & 0.0934 & 0.0947 & 0.1188  & 0.2674&0.2565 &0.3615\\ 
& &\textbf{ER$_{can}$} & 0.0644& 0.0669&0.0785  & 0.1951& 0.1935&0.2725\\ 
& &\textbf{ER$_{doc}$} & 0.0558&0.0582 &0.0678  & 0.1851& 0.1832&0.2598\\ 
\midrule

& &\textbf{DR$_{sparse}$} & 0.1021 & 0.1037 & 0.1306   & 0.2837&0.2739 &0.3784\\ 
& &\textbf{DR$_{flat}$} & 0.1058 & 0.1074 & 0.1338   & 0.2902& 0.2771&0.3895\\ 
& &\textbf{DR$_{hnswpq}$} & 0.0941 & 0.0911 & 0.1258   & 0.2912&0.2718 &0.3933\\ 
\textbf{Quote} &  \textbf{Summary} & \textbf{DR$_{hnsw}$} & 0.1038 & 0.1056 & 0.1309  &0.2882 & 0.2756&0.3866\\ 
& &\textbf{DR$_{pq}$} & 0.1023 & 0.1013 & 0.1293  & 0.2882&0.2750 &0.3854\\ 
& &\textbf{ER$_{can}$}  & 0.0568&0.0585 &0.0717  & 0.2002&0.1958 &0.2778\\ 
& &\textbf{ER$_{doc}$}  &0.0457 & 0.0473& 0.0569 &0.1850 & 0.1799&0.2627\\ 
\toprule

\end{tabular}}
\caption{\textbf{DR} denotes the document retrieval approach, and the subscripts represent 5 types of retrieval indices mentioned in Section \ref{ERmethods} Approach 1, Lucene sparse bag-of-words index, Faiss flat index, Faiss HNSWPQ index, Faiss HNSW index, and Faiss PQ index. \textbf{ER}$_{can}$ is the candidate-based expert finding approach, and \textbf{ER}$_{doc}$ is the document-based expert finding approach. In two expert finding approaches the input query length is set to 5.}
\label{table:results}
\end{table*}

\section{Noisy Sample}
Five types of noise and their corresponding examples that appeared in the raw test set are listed in Table \ref{tab:noisy}.

\section{Results: Expert Recommendation}
\label{sec:appendix-results}
Table \ref{table:results} shows the experimental results of expert recommendation. Among all 6 document-query combinations, the document retrieval (\textbf{DR}) approaches outperform the expert retrieval (\textbf{ER}) approaches. Among various document indexing strategies, the Lucene sparse bag-of-words index (\textbf{DR$_{sparse}$}) gives better results compared to other dense transformer-encoded Faiss indices. Averagely, experiments perform better when we use the context as our documents. We believe that adjacent contexts can also contain information about the sources and their quotations. 

\end{document}